\documentstyle[preprint,aps,prb]{revtex}
\def\ibd {{\it ibid. }}
\def\bfq{{\bf q}}

\def\bfq'{{\bf q'}}
\begin{document}
\preprint{YUMS-98-14/SNUTP-98-090}
\title{Fluctuation correction to the ground state energy density of a dilute Bose gas
in the functional Schr\"{o}dinger picture}
\author{Sang-Hoon Kim,$^1$ Hyun Sik Noh,$^2$ Dae Kwan Kim,$^3$ Chul Koo Kim,$^3$
   and Kyun Nahm,$^4$}
\address{$^1$Division of Liberal Arts, Mokpo National Maritime University,
 Mokpo 530-729, Korea}
\address{$^2$Hanlyo Sanup University,
Kwang-Yang 545-800, Korea}
\address{$^3$Department of Physics and Institute for Mathematical Sciences, Yonsei University,
         Seoul 120-749, Korea}
\address{$^4$Department of Physics, Yonsei University, Wonju 220-710, Korea}
\date{\today}
\maketitle
\begin{abstract}
A dilute Bose gas system is studied using the functional Schr\"{o}dinger  picture theory.
The ground state properties are obtained by solving the infinite dimensional  Schr\"{o}dinger
 equation variationally. It is shown that a shifted Gaussian trial  wavefunctional enables us
 to calculate a higher order correction, which corresponds to the fluctuation contribution
 from the condensate.  The obtained term is compared with the quantum correction arising from
 the low energy $3 \rightarrow 3$ scattering.
\end{abstract}
\draft
\newpage
\section{Introduction}
Recent successful demonstrations of Bose-Einstein condensation\cite{ande,brad,davi}
 have spurred a revival of interest in Bose gases.  In order to gain a deeper
 understanding than the mean field result.\cite{baym} it is necessary to
 calculate correction from quantum fluctuations around the mean field.
The problem of interacting Bose gas has a long history.  The standard result
for the ground state energy density given in standard many-body
textbooks\cite{abri,fett,maha} has the form
\begin{equation}
\frac{E_g}{V} = \frac{{2 \pi  \hbar^{2} a n^2 }}{ m}
 \left\{ 1+ \frac{128}{15\sqrt{\pi} } (n a^3)^{1/2} +
\left[ \frac{8(4\pi - 3\sqrt{3})}{3} \ln(n a^3) + \kappa \right]
n a^3 + ...  \right\},
\label{1}
\end{equation}
where $n=N/V$ is the number density of atoms and $a$ is the
S-wave scattering length. Although the coefficient of $\ln(na^3)$ in the $na^3$
correction was calculated early in 1950's,\cite{wu,huge,sawa} a successful
calculation of the constant $\kappa$ had to wait until recently.
Braaten and Nieto\cite{braa} employed an effective field theory and the renormalization
scheme to obtain the value of $\kappa$ in terms of $a$ and atomic radius $R_0$.
In this paper, we show that the interaction of particles out of or into the
condensate give also an $na^3$ contribution to the ground state energy in
addition to the $3 \rightarrow 3$ scattering contribution of Braaten and
Nieto.\cite{braa}  The relative importance of each term will be
discussed.

In the following, we apply the functional Schr\"{o}dinger  picture(FSP) scheme
to a non-relativistic Bose gas to obtain the ground state properties.
The Schr\"{o}dinger  picture for many particle systems or relativistic quantum
field theory is an infinite dimensional extension of quantum mechanics and
leads into a functional formulation.\cite{corn}
The FSP method in the relativistic quantum field theory is known to have
several advantages over the conventional Green's function
method.\cite{kim,stev}
One of such advantages is that the FSP method can readily be combined with a
variational method to yield non-perturvative results.

The FSP formulation of many-body fermionic system has been shown to yield the
Hartree-Fock results when coupled with a Gaussian trial
functional.\cite{noh1,noh2}
In order to obtain a result going beyond the mean field value, we employ a
shifted Gaussian trial functional and calculate a contribution arising from the
condensation fluctuation.

\section{Many-body dilute Bose gas}
We start with the standard model Hamiltonian for an interacting Bose
gas\cite{abri,fett,maha,bogo}
\begin{equation}
\hat{H} = \sum_{ k }  \frac{\hbar^2 k^{2}}{2m} a_{k}^{\dag} a_{k}
+ \frac{U}{2V} \sum_{ {{ k_{1} k_{2}}  \atop {k_{3} k_{4} }}  }
  a_{k_{1}}^{\dag}  a_{k_{2}}^{\dag} a_{k_{3}} a_{k_{4}}
  \delta_{k_{1}+k_{2},k_{3}+k_{4}},
\label{2}
\end{equation}
Here, the constant  matrix element  $U$ should  be determined  by requiring that
$H$ produces the two-body scattering  properties in vacuum.
Since the system is a dilute gas, we assume that the scattering length
  $a$ is much less than the inter-particle spacing $ n^{1/3}$.
Therefore, the validity of the present calculation is restricted by the condition
$n a^3 \ll 1$.  Under this  assumption, the  zero momentum  operators
$a_{0},  a_{0}^{\dag}$  become nearly classical due to occurrence
 of a condensate.  The interacting part of the Hamiltonian is
 rewritten explicitly in terms of $a_{0}, a_{0}^{\dag} $ as
\begin{eqnarray}
{\hat{H}}_{int} &=& \frac{U}{2V} a_{0}^{\dag} a_{0}^{\dag}
a_{0} a_{0}
\nonumber \\
&+& \frac{U}{2V} \sum_{ k \neq 0 }
 \left[ 2( a_{k}^{\dag} a_{0}^{\dag} a_{k} a_{0}
+ a_{-k}^{\dag} a_{0}^{\dag} a_{-k} a_{0} )
+ a_{k}^{\dag} a_{-k}^{\dag} a_{0} a_{0}
+ a_{0}^{\dag} a_{0}^{\dag} a_{k} a_{-k}  \right]
\nonumber \\
  &+& \frac{U}{V} \sum_{ k \neq q \neq  0 }
 \left[   a_{k+q}^{\dag} a_{0}^{\dag}  a_{k} a_{q}
+ a_{k+q}^{\dag} a_{-q}^{\dag} a_{k} a_{0}  \right],
\label{3}
\end{eqnarray}
where only terms of order of $N_{0}^{2}, N_{0}$, and $\sqrt{N_0}$
have been retained.  Three zero momentum operators term does
 not exists, because any selection of three zero wave-vectors
make the fourth zero by the momentum conservation.
It should be noted that the last term arises from the interactions of particles
out of and into the condensate, thus representing the condensation
fluctuation.\cite{fett}

In order to study the contribution from the last term,
we  introduce a new  variable $\gamma_{k}$
\begin{equation}
\gamma_{k} = \sum_{q \neq 0}  a_{k+q}^{\dag} a_{q}.
\label{4}
\end{equation}
Also, we take the mean field approximation  of
$\gamma_k$ as $ \sum_{q \neq 0} \langle a_{k+q}^{\dag} a_{q}\rangle$.
Replacing the operators $a_{0} , a_{0}^{\dag}$  by  $\sqrt{N_0}$,
 ${\hat{H}}_{int}$ approximately becomes
\begin{eqnarray}
{\hat{H}}_{int} &=& \frac{U }{2V}N_{0}^{2}
\nonumber \\
 &+& \frac{U }{2 V}N_{0} \sum_{k \neq  0} \left[
 2 (a_{k}^{\dag} a_{k}+a_{-k}^{\dag} a_{-k}) +
(a_{k}^{\dag} a_{-k}^{\dag} + a_{k} a_{-k}) \right]
\nonumber \\
&+& \frac{U }{V}\sqrt{N_0} \sum_{k  \neq  0}
 \gamma_{k} ( a_{k}+ a_{-k}^{\dag} ),
\label{5}
\end{eqnarray}

The number operator satisfies the relation
\begin{equation}
N_0  = \hat{N} - \frac{1}{2}\sum_{k \neq   0}
( a_{k}^{\dag} a_{k} +  a_{-k}^{\dag} a_{-k}).
\label{6}
\end{equation}
Using this relation and keeping up to the
 terms of $\sqrt{N}$, the model Hamiltonian  becomes
\begin{eqnarray}
\hat{H} &=&  \frac{1}{2} n^2 V U
 \nonumber \\
  &+& \frac{1}{2} \sum_{k \neq  0}
 \left[ (\epsilon^{0}_k + n U) (a_{k}^{\dag} a_{k} + a_{-k}^{\dag} a_{-k})
 + n U (a_{k}^{\dag} a_{-k}^{\dag} + a_{k} a_{-k})\right]
 \nonumber \\
 &+& \frac{nU}{\sqrt{N}}
 \sum_{k \neq  0} \gamma_k ( a_k  + a_{-k}^{\dag}  ),
\label{7}
\end{eqnarray}
where $\epsilon^{0}_k = \hbar^2 k^{2}/2m$.
The ground state properties of the above Hamiltonian is to be calculated using
a variational scheme of the FSP.

\section{Functional Scr\"{o}dinger picture method of Bosonic systems}
The FSP approach to many-boson system is simpler than many-fermion
systems\cite{corn,hatf} when the Bose-Einstein condensation(BEC) is not
included.  However, when the BEC exists, a special care is necessary to include
the condensation in the theory.
In the FSP theory, the bosonic field operators satisfy the equal time commutation  relation
\begin{equation}
 \left[ \Psi (x), \Psi^{\dag}(y) \right] = \delta(x-y) .
\label{8}
\end{equation}

Since we are working on a homogeneous system, it is convenient to work in the
$k$-space;
\begin{eqnarray}
 \Psi (x) &=&  \int \frac{d^{3}k}{(2\pi)^{3/2} }
\, a_k \, e^{i \vec{k} \cdot \vec{x}},
\nonumber \\
 \Psi^{\dag} (x)  &=& \int \frac{d^{3}k}{(2\pi)^{3/2}}
\, a_{k}^{\dag} \, e^{-i\vec{k} \cdot \vec{x}}.
\label{9}
\end{eqnarray}
We find that the fundamental commutation relations in Eq. (\ref{8}) is  satisfied,
if $a_k$ and $a_k^\dag$ are expressed as functional operators as follows,
\begin{eqnarray}
a_{k} & \rightarrow &   \alpha_k \phi(k),
\nonumber \\
a_{k}^{\dag}  &\rightarrow & - \alpha_k  \frac{\delta}{\delta
\phi(k)}.
\label{11}
\end{eqnarray}
The weighting factor $\alpha_k$ is necessary to correctly take account the BEC.
$\alpha_k$ is assumed real and even under  under $ k \rightarrow -k$.
The above expressions and the commutation relation,
$\left[\phi(k),-\frac{\delta}{\delta\phi(k')}\right] =
 \delta(k-k')$, yield the constraint
 \begin{equation}
 \alpha_k\alpha_{k'}\delta(k-k') = \delta_{k k'}.
 \label{112}
 \end{equation}
Note that $\alpha_k^2\delta(0) = 1$   except $ k=0$.
 This is true only when $\alpha_k $ is almost zero-flat
for any $k$ except $k=0$.

The Schr\"{o}dinger equation in $k$ space is expressed as
\begin{equation}
\hat{H} \left( \phi(k), -\frac{\delta}{\delta \phi(k)} \right) \Phi [\phi]
= E  \Phi [\phi].
\label{14}
\end{equation}
In the next section, we solve this functional
differential equation using a variational method.

\section{The ground state energy density}
The explicit form of the Hamiltonian in  Eq. (\ref{14})
 is obtained by substituting Eq. (\ref{11}) into Eq. (\ref{7}),
 \begin{eqnarray}
 \hat{H} \left( \phi (k), -\frac{\delta}{\delta \phi(k)} \right)
  =  \frac{1}{2} n^{2} V U
\nonumber \\
 +   \frac{1}{2}
  \sum_{ k \neq  0 }\alpha^2_k\left\{ - ( \epsilon^{0}_k +n U)
 \left[ \frac{\delta}{\delta \phi(k)} \phi(k)
 +\frac{ \delta }{ \delta \phi(-k)} \phi(-k) \right] \right.
 \nonumber  \\
     \left. + \, n U \left[ \frac{\delta}{\delta  \phi(k)}
 \frac{\delta}{\delta \phi(-k)} +\phi(k)\phi(-k) \right]\right\}
\nonumber \\
 + \frac{nU}{\sqrt{N}}\sum_{ k \neq  0 }
\alpha_k \gamma_k
 \left[ \phi(k)- \frac{\delta}{\delta  \phi(-k)}
\right].
 \label{15}
 \end{eqnarray}

Since we are looking for the ground state,  we may assume that  the ground
state functional is real,  positive, and has no nodes anywhere.\cite{penr}
Thus, we write
\begin{equation}
\Phi_{0} [ \phi ] = C e^{ G[\phi] } ,
\label{16}
\end{equation}
where $C$ is a normalization constant.
  When this expression is substituted into  in Eqs. (\ref{14}) and (\ref{15}),
we obtain the relation
\begin{eqnarray}
  \frac{1}{2} n^{2} V U
\nonumber \\
- \frac{1}{2} \sum_{k \neq 0}\alpha^2_k (\epsilon^{0}_k +n U)
  \left[ 2\delta(0)  + \phi(k) \frac{ \delta G}{\delta \phi(k)}
+ \phi(-k) \frac{\delta G}{\delta \phi(-k)} \right]
\nonumber  \\
 + \frac{n U}{2} \sum_{k \neq 0}\alpha^2_k \left[ \frac{\delta^2 G}
 {\delta \phi(k) \delta \phi(-k)}
+ \frac{\delta G}{\delta \phi(k)} \frac{\delta G}{\delta \phi(-k)}
+ \phi(k) \phi(-k) \right]
\nonumber \\
  +  \frac{nU}{\sqrt{ N}}  \sum_{k \neq 0}
\alpha_k \gamma_k
\left[ \phi(k) -  \frac{\delta G}{\delta \phi(-k)}
 \right] = E_{g}.
 \label{17}
\end{eqnarray}

The power  counting  technique\cite{hatf}  tells  us that
 $ G[ \phi ] $ is minimally  quadratic in $ \phi (k) $.
Thus, one can write $G[ \phi ]$ as a  shifted Gaussian in $\phi$,
\begin{equation}
G[ \phi ] = \int d^{3} k \left[ g(k) \phi (k) \phi (- k)
+ f(k) \phi (k) \right],
\label{171}
\end{equation}
where $g(k)$ and $f(k)$ are variational function parameters
 that will be determined in the below.
The shifted Gaussian form of the trial function
for the ground state allows us to calculate the contributions from
 linear terms of field operators.  The trial functional method does not
work if the Hamiltonian contains  triplet field operator terms.
This explains why we had to introduce the mean field operator $\gamma_k$
in the Hamiltonian.

 Substituting Eq. (\ref{171}) into Eq. (\ref{17}), we obtain the following
 relation,
\begin{eqnarray}
  \frac{1}{2} n^{2}  V U
  + \sum_{ k \neq 0 } \alpha^2_k\delta(0)
 \left[ n U g(k) -  ( \epsilon^{0}_k +n U )\right]
\nonumber \\
 + \frac{nU}{2} \sum_{ k \neq 0 }\left[\alpha_k^2 f(k)f(-k)
 - \frac{2\alpha_k\gamma_{-k}}{\sqrt{N}}  f(k) \right]
\nonumber \\
 + \sum_{k \neq 0}\alpha^2_k
\left[ -(\epsilon^{0}_k +n U )f(k) + 2nU g(k) f(k)
+ \frac{nU\gamma_k}{\alpha_k\sqrt{ N}}\left\{1-2g(k)\right\}
 \right]\phi(k)
\nonumber \\
+ \frac{1}{2}  \sum_{k \neq 0}\alpha^2_k
 \left[ - 4(\epsilon^{0}_k +n U ) g(k)
  + 4 n U g^{2} (k)+  n U \right] \phi(k) \phi(-k)= E_{g} .
 \label{20}
\end{eqnarray}
        Counting powers of both sides, we obtain
\begin{equation}
 4 n U g^{2} (k)- 4 ( \epsilon^{0}_k +n U ) g(k) +  n U =0,
\label{21}
\end{equation}
\begin{equation}
 -(\epsilon^{0}_k +n U )f(k) + 2nU g(k) f(k)
+ \frac{nU\gamma_k}{\alpha_k\sqrt{ N}}\left[1-2g(k)\right]
 = 0,
\label{211}
\end{equation}
and
\begin{eqnarray}
E_{g} =  \frac{1}{2} n^{2} V U
+ \sum_{k\neq 0}\left[n U g(k)-(\epsilon^{0}_k+nU)\right]
\nonumber\\
 +\frac{nU}{2}\sum_{ k \neq 0 } \left[ \alpha^2_k f(k) f(-k)
- \frac{2\alpha_k\gamma_{-k} }{\sqrt{N}}f(k) \right].
\label{22}
\end{eqnarray}
The constraint, Eq. (\ref{112}), was used above.

 Eq. (\ref{21}) for the variational
function $g(k)$ can be easily solved to yield that
\begin{equation}
g(k)= \frac{ \epsilon^{0}_k +n U + E(k) }{2 n U },
\label{410}
\end{equation}
and
\begin{equation}
f(k)= \frac{\gamma_k}{\alpha_k\sqrt{ N}}
 \frac{  \epsilon^{0}_k  + E(k)} {E(k)},
\label{411}
\end{equation}
where
\begin{equation}
E(k) = \sqrt{ ( \epsilon^{0}_k +n U )^{2} - (n U )^{2} }.
\label{412}
\end{equation}
$E(k)$ represents the excitation energy spectrum.

The expressions of $g(k)$ and $f(k)$ allow us to evaluate
the ground state energy.
\begin{eqnarray}
 E_g  &=& E_0 + E_1 + E_2
 \nonumber \\
 &=& \frac{1}{2} n^{2} V U - \frac{1}{2} \sum_{ k \neq  0 }
 \left[ (\epsilon^{0}_k + n U) - E(k) \right]
- \frac{n U^2}{V}\sum_{ k \neq  0 }\frac{\gamma_k \gamma_{-k}}
{ \epsilon^{0}_k +2nU}.
\label{25}
\end{eqnarray}
The first two terms, $E_0$ and $E_1$ in the ground state energy
are exactly  same as those obtained through  other methods
in  the  literature.\cite{abri,fett,maha}
Following the standard step relating the scattering length to the interaction,
 \begin{equation}
 \frac{4\pi \hbar^2 a}{m}
 = U - \frac{U^2}{2V}\sum_{k \neq 0}\frac{1}{\epsilon^{0}_k} + ...\, ,
 \label{416}
 \end{equation}
we obtain  the ground state energy density of the first two terms as
\begin{equation}
\frac{E_0 + E_1}{ V}
 = \frac{{2 \pi  \hbar^{2} a n^2 }}{ m} \left[ 1+ \frac{128}{15\sqrt{\pi}}
 (na^3)^{1/2}  \right].
\label{261}
\end{equation}
The total ground state energy in the present calculation
have an additional contribution $E_2$,
which will be calculated in the following sections.
The corresponding wavefunctional is given by
\begin{equation}
\Phi_{0} [ \phi ] = C  \exp \left\{
 \int d^{3} k \left[ \frac{\epsilon^{0}_k +n U  + E(k)} {2 n U } \phi(k)\phi(-k)
+\frac{\gamma_k}{\alpha_k\sqrt{ N}} \frac{  \epsilon^{0}_k  + E(k)}
 {E(k)} \phi(k)\right] \right\}.
\label{27}
\end{equation}
$\gamma_k$ and $\alpha_k$ are discussed in the following sections.

\section{The particle depletion}
 The particle depletion from the zero momentum condensate is
 defined by
\begin{equation}
\frac{N-N_0}{N}
= \frac{1}{N}\sum_{k \neq 0} \langle a_k^{\dag} a_k\rangle.
\label{61}
\end{equation}
This value is  equal to $\gamma_0/N$ in Eq. (\ref{4}),
 and  can be calculated using Eqs. (\ref{11}), (\ref{16}), and (\ref{171}).
We obtain for $n_k = <a_k^\dag a_k>$,
\begin{eqnarray}
 n_k  &=& \langle -\alpha_k^2 \left[ \delta(0)
+ \phi(k)\frac{\delta G}{\delta \phi(k)}\right]\rangle
\nonumber\\
&=& \langle -1 + 2\alpha_k^2 |g(k)| \phi^2(k)-\alpha_k^2f(k)\phi(k)\rangle.
\label{63}
\end{eqnarray}
The average value of $\phi^2(k)$ and $\phi(k)$
 in $k$ state are also obtained as
 \begin{equation}
  \frac{\langle \Phi_0 | \phi^2(k) | \Phi_0 \rangle}
 {\langle \Phi_0 | \Phi_0 \rangle}
 =  \frac{1}{4|g(k)|} + \frac{f^2(k)}{4|g(k)|^2},
\label{64}
\end{equation}
and
\begin{equation}
  \frac{\langle \Phi_0 | \phi(k) | \Phi_0 \rangle}
 {\langle \Phi_0 | \Phi_0 \rangle}
 =   \frac{f(k)}{2|g(k)|}.
\label{65}
\end{equation}
Substituting the above results into Eq. (\ref{63}),
we obtain the average number of particle in $k$ state as
 \begin{equation}
 n_k = - 1 + \frac{\alpha_k^2}{2},
 \label{66}
\end{equation}
where  $k =0$ is not included. Clearly, we observe that $\alpha_k$ cannot be 1
in the FSP formalism with the BEC.  At the same time,  $n_k$
should satisfy the following two conditions:
\begin{equation}
\lim_{k\rightarrow 0} n_k = \infty,
\,\, \mbox{and} \,\,
\lim_{k\rightarrow \infty} n_k = 0.
\label{68}
\end{equation}

We now use a simple comparison method to calculate the particle depletion.
We have only two free variables of $k$, $g(k)$ and $E(k)$,
in the ground state wavefunctional.
 $f(k)$ does not contribute to the average particle depletion but represents the
 fluctuation of the condensate.        From Eqs. (\ref{410}) - (\ref{412})
 we obtain the following properties:\\
(i) As $ k \rightarrow 0,$
\begin{equation}
g(0)=\frac{1}{2},  \,\,  \mbox{and} \,\,   E(0)= 0.
\label{241}
\end{equation}
(ii) As $ k \rightarrow \infty,$
\begin{equation}
n U g(k) \, \sim \, E(k) \,\,\, (\,\, \rightarrow \epsilon^{0}_k).
\label{242}
\end{equation}
Using the above properties, we observe that
 the simplest way to satisfy these conditions is
\begin{equation}
\frac{\alpha_k^2}{2} = \frac{n U g(k)}{E(k)}.
\label{69}
\end{equation}
Indeed, $\alpha_k$  has the desired properties of a very sharp
 peak at $k=0$, and almost zero values elsewhere.

The particle depletion is now calculated as
\begin{eqnarray}
\frac{N-N_0}{N}
&=&\frac{1}{N} \sum_{k\neq 0}
\left[ \frac{n U g(k)}{E(k)} -1 \right]
\nonumber\\
&=& \frac{1}{2N}\sum_{k\neq 0}
\left[\frac{\epsilon^{0}_k + nU}{E(k)} - 1 \right].
\label{71}
\end{eqnarray}
  The the summation over $k$ and substitution of the S-wave
 scattering length $a$ yield the standard result\cite{abri,fett,maha}
\begin{equation}
\frac{N-N_0}{N}=\frac{8}{3}
\left( \frac{na^3}{\pi}\right)^{1/2}.
\label{72}
\end{equation}
So far, we have shown that the FSP approach coupled with
 the variational method can produce the standard results on the ground state
 energy and the particle depletion in a straightforward manner.
In the following, we discuss a higher energy correction.

\section{The fluctuation correction to the ground state energy density}

The higher order contribution to the ground state energy
in Eq. (\ref{25}) comes from the interactions of particles out of and into the condensate.
Clearly, this term represents the fluctuation contribution from the condensate
and disappears if the number of condensate particles are conserved strictly.
The fluctuation contribution, $E_2$, carries the ultraviolet divergence.  The
same divergence in $E_1$ was handled through the expression of the scattering
length $a$ to order of $U^2$ as given in  Eq. (\ref{416}).
However, for the calculation of $E_2$ such a simple cancellation scheme is not possible,
because it is now necessary to expand $a$ up to $U^3$.
Therefore, we resort to an altanative cutoff procedure,
called ``minimal subtraction".\cite{braa}
In minimal subtraction, linear, quadratic, and other power ultraviolet
divergences are removed as part of the regularization scheme by subtracting the
appropriate power of $k$ from the momentum space integrated.
Following this prescription, we obtain $E_2$ in a dimensionless form
\begin{equation}
\frac{E_2}{E_c} = -
 \frac{2U}{n V^2} \sum_{k \neq 0}\gamma_k \gamma_{-k}
\left[\frac{1}{\epsilon_k^0 + 2 n U} - \frac{1}{\epsilon_k^0} \right],
\label{81}
\end{equation}
where $E_c = 2\pi\hbar^2 a n^2 V/m$.

Now, it is necessary to evaluate
$\gamma_k \gamma_{-k} = \sum_{q,q' \neq 0}
 a_{k+q}^\dagger a_q a_{-k+q'}^\dagger a_{q'}$.
 It is known that a dominant contribution arises when a particle
 interacts with itself, which happens when
  ${\bf q'} = {\bf k} + {\bf q}$.\cite{maha}  Therefore,
\begin{eqnarray}
 \gamma_k \gamma_{-k} &\simeq& \sum_{q \neq 0}
 a_{k+q}^\dagger a_{k+q} a_{q}^\dagger a_q
\nonumber \\
&\simeq&  \sum_{q \neq 0}  n_q^2
\nonumber \\
&=& N\left(\pi-\frac{8}{3}\right)\sqrt{\frac{n a^3}{\pi}}.
\label{83}
\end{eqnarray}
In the second step, we used the fact that $n_0$ is
much larger than $n_k (k\neq 0)$.
For the last step, Eqs. (\ref{66}) and (\ref{69}) are used.
This approximation gives a simple result for the
second order correction, $E_2$,
\begin{equation}
\frac{E_2}{E_c}
=  16\left(\pi-\frac{8}{3}\right) n a^3.
\label{84}
\end{equation}
This fluctuation correction has the same $n a^3$ dependence with the
$3\rightarrow 3$ scattering contribution obtained by
Braaten and Nieto.\cite{braa}  This point will be discussed in detail later
in this section.

In the present approximation scheme, the  total
ground state energy density is given by
\begin{equation}
\frac{E_g}{V} = \frac{{2 \pi  \hbar^{2} a n^2 }}{ m}
 \left [ 1+ \frac{128}{15\sqrt{\pi} }
 (na^3)^{1/2} +16 \left(\pi-\frac{8}{3}\right) n a^3 \right ].
\label{85}
\end{equation}
And the corresponding chemical potential is given by
\begin{eqnarray}
\mu &=& \frac{\partial E_g}{\partial N}
\nonumber\\
&=& \frac{4\pi \hbar^2 a n}{m} \left[ 1+ \frac{32}{3\sqrt{\pi}}
(na^3)^{1/2}
+24\left(\pi-\frac{8}{3}\right) n a^3\right].
\label{86}
\end{eqnarray}

It was shown that the low-density expansion for the
energy density has the general form,\cite{huge,braa}
\begin{equation}
\frac{E_g}{V} =\frac{2\pi\hbar^2a n^2}{m}\sum_{n=0}^\infty
\sum_{l=0}^{[n/2]} C_{nl} (na^3)^{n/2} \ln^l(n a^3),
\label{120}
\end{equation}
where $[n/2]$ is the Gauss number which takes the integer part of $n/2$.
Comparing this equation with Eq. (\ref{1}), we find the standard results
 $C_{00}=1$, $C_{10}= 128/15\sqrt{\pi}$, and
  $C_{21}= 8(4\pi - 3\sqrt{3})/3$.
In addition, Braaten and Nieto calculated the constant under logarithm, which
corresponds to $\kappa_1 = \frac{8(4\pi - 3\sqrt{3})}{3} \ln(2.8  R_0^2/ a^2)$,
when the scattering length $a$ is much larger than the atomic radius $R_0$.
Since the result, $\kappa_2$, we obtained within the approximation of a shifted
Gaussian functional does not depend on the atomic parameters and has a
different physical origin, we believe that the two constants are independent
and additive so that $\kappa = \kappa_1 + \kappa_2$.
Comparing the two terms, we observe that $\kappa_2$ is dominant if $R_0/a <
0.75$.  In the real alkali atom gases, generally $R_0/a \sim 0.1$.\cite{nist}
Therefore, we believe that  $\kappa_2$ is a larger contribution in general.

\section{summary}

We have presented a functional Schr\"{o}dinger picture approach to a dilute
Bose gas system.  Employing a shifted Gaussian trial wavefunctional,
we have calculated the ground state energy density and the particle depletion.
It is shown that the present scheme allows to calculate the ground state
properties in a straightforward manner and, in addition, to obtain the
fluctuation contribution beyond the mean field result.
The obtained correction is compared with the quantum correction term obtained
through the low energy $3\rightarrow 3$ scattering.

In  the functional Schr\"{o}dinger picture scheme,  the many-body
Schr\"{o}dinger equation is solved directly without resorting to the Green's
function or a canonical transformation.
Therefore, this scheme provides an intuitive alternative to conventional
many-particle physics.

 \acknowledgements
 This work was has been partly supported by the Korea Ministry
 of Education (BSRI-97-2425)
 and the Korea Science and Engineering Foundation through project No.
 95-0701-04-01-3 and also through the SRC program of SNU-CTP.


\end{document}